\documentclass[aps,prl,groupedaddress,superscriptaddress,twocolumn]{revtex4}

\usepackage{graphicx}

\begin{document}

\title{Band-like transport and trapping in TMTSF Single Crystal Transistors}%

\author{H. Xie}
\affiliation{ Kavli Institute of NanoScience, Delft University of
Technology, the Netherlands}
\affiliation{DPMC and GAP, University of Geneva, 24 quai Ernest-Ansermet, CH1211 Geneva, Switzerland}
\author{H. Alves}
\affiliation{ INESC-MN and IN, Rua Alves Redol 9, 1000-029 Lisboa,
Portugal}
\author{A. F. Morpurgo}
\email{Alberto.Morpurgo@unige.ch}
\affiliation{DPMC and GAP, University of Geneva, 24 quai Ernest-Ansermet, CH1211 Geneva, Switzerland}

\iffalse
\author{$H. Xie^{1,2}$, $ H. Alves^{1}$, $A. F. Morpurgo^{2}$ }%
\email{Alberto.Morpurgo@unige.ch} \affiliation \\
$^{2}$ DPMC and GAP, University of Geneva, 24 quai Ernest-Ansermet, CH1211 Geneva, Switzerland}
\date{Feburary 26, 2009}%
\fi

\begin{abstract}
We perform a combined experimental and theoretical study of tetramethyltetraselenafulvalene (TMTSF) single-crystal field-effect transistors, whose electrical characteristics exhibit clear signatures of the intrinsic transport properties of the material. We introduce a simple, well-defined model based on physical parameters and we successfully
reproduce quantitatively the device properties as a function of temperature and carrier density. The analysis allows its internal consistency to be checked, and enables the reliable extraction of the density and characteristic energy of shallow and deep traps in the material. Our findings indicate that shallow traps originate from electrostatic potential fluctuations generated by charges fixed in the deep traps.
\end{abstract}
 \maketitle

Despite impressive progress in the use of organic semiconductors for the realization of practical circuits and
devices, our fundamental understanding of these materials remains limited. In this context, the recent observation  of an anisotropic carrier mobility increasing with lowering temperature in organic single-crystal field-effect transistors (FETs)\cite{Review} establishes a breakthrough, because it opens the possibility to investigate intrinsic transport properties of molecular materials at finite and tunable carrier density\cite{Sundar04, Podzorov04}.
Surprisingly, a systematic, in-depth analysis of the intrinsic transport phenomena observed in single-crystal FETs has not been attempted yet. In large part, this is due to the lack of a well-defined model enabling a
quantitative study to be performed. In fact, past work on the
intrinsic transport properties of organic semiconductors --based on
time-of-flight measurements-- has relied on the so-called
trap-and-release model\cite{Karl, Horowitz}, which, despite being helpful\cite{Calhoun},
cannot be applied to the systematic analysis of systems at finite carrier density. This poses
a clear problem, because a quantitative analysis of the properties
of organic semiconductors is essential to characterize devices and
materials unambiguously, to perform comparative studies, and to reach
a true microscopic understanding, which is essential for the efficient long-term development
of practical applications.

We address this issue through a combined experimental and theoretical study of transport through tetramethyltetraselenafulvalene (TMTSF, C$_{10}$H$_{12}$Se$_{4}$, see Fig. 1a) single-crystal FETs. We first show that these devices exhibit the  characteristic signature of the intrinsic transport properties. We then introduce a simple model based on physically transparent quantities and assumptions, with which we analyze in detail the temperature and density dependence of the mobility, and the temperature dependence of the threshold voltage. Our analysis reproduces the observations systematically and quantitatively, it gives consistent results on different devices,  it enables the reliable determination of the concentration and characteristic energy of traps, it allows the check of its internal consistency to be performed, and it points to the physical origin of the shallow traps. These results clearly indicate
that the model represents an appropriate framework for the analysis through high-quality organic FETs.

So far, rubrene crystals have been the only ones to unambiguously
show the expected signatures of intrinsic transport at finite
carrier density in a FET configuration\cite{Review}. We have
explored crystals of different molecules, and here we discuss TMTSF
transistors\cite{Chaikin}, realized by laminating a single-crystal
grown from vapor phase transport onto a polydimethylsiloxane (PDMS)
support as shown in Fig. 1b,c (the fabrication procedure is
identical to that described in  Ref. \cite{Menard}). Immediately
after fabrication, the devices were transferred into the vacuum
chamber ($p\approx 2 \ 10^{-7}$ mbar) of a flow cryostat where all
transport measurements were done. Most measurements  were made using
a HP 4156A parameter analyzer, in a four-terminal configuration
(Fig. 1 (c)), to eliminate the effects of the contact resistance. We
discuss data taken on three different FETs, representative of the
approximately 10 devices studied.

\begin{figure}[htbp]
\includegraphics[width=0.8\linewidth]{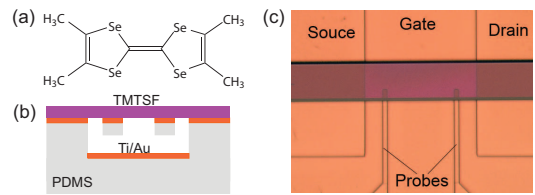}
  \caption {(a) Chemical structure of TMTSF molecule. (b) Schematic cross section of our PDMS-based
  TMTSF transistors. (c) Optical microscope image of a TMTSF PDMS FET (the seperation between source and drain is 300 $\mu$m). }
  \label{FIG. 1}
\end{figure}

Fig. 2(a) shows the source-drain current (I$_{SD}$) measured at
different temperatures, as a function of gate voltage (V$_{G}$), and
the inset $I_{SD}$ measured on the same device as a function of
$V_{SD}$ at T=300 K, for different values of $V_G$. The measurements
exhibit the characteristic behavior of high-quality devices (e.g.,
absence of hysteresis, stability and reproducibility). The charge
carrier mobility is extracted from the linear part of
I$_{SD}$-V$_{G}$ characteristics, using the relation
$\mu=\frac{L}{W} \frac{1}{C_i}\frac{1}{V}\frac{dI_{SD}}{dV_G}$ (with
$L$ separation between voltage probes, $W$ crystal width, C$_{i}$
gate capacitance per unit area, and V voltage difference measured
between the two voltage probes). At room temperature, $\mu = 4$
cm$^{2}$/Vs reproducibly (the  same value is obtained from the
$I_{SD}-V_{SD}$ characteristics in the saturation regime). From the
$I_{SD}-V_G$ curves we extract the full temperature dependence of
the mobility, as well as the threshold voltage $V_{T}(T)$ (by
extrapolating the curves to zero current).

\begin{figure}[htbp]
\includegraphics[width=0.8\linewidth]{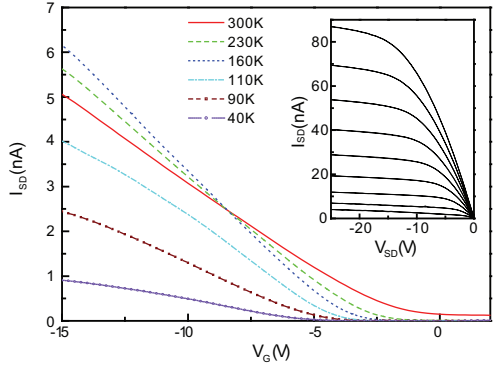}
  \caption { Source-drain current, measured as a function of gate voltage at different temperatures
  ($V_{SD}$=1 V). The inset shows the source-drain current as a function of source-bias voltage, for different values of $V_G$ (from -1V to -19V, in -2 V steps). }
  \label{FIG. 2}
\end{figure}

Fig. 3a and b show the mobility and threshold voltage as a function
of temperature measured on three different samples. The mobility
tends to increase, exhibiting an identical temperature dependence
down to a sample-dependent temperature value, after which $\mu$
decreases with further lowering $T$. The temperature interval where
$\mu(T)$ is "metallic-like" reflects the intrinsic material
properties, and the different extents of this interval in the three samples
indicate that different amount of disorder is present.
Interestingly, the decrease of the mobility at low $T$ is very slow.
In our best sample, $\mu$ reaches a value above 6 cm$^2$/Vs at
$T\simeq 160$ K, and it is still close to 2 cm$^{2}$/Vs at 50 K. The
behavior  of the threshold voltage is qualitatively similar in
different devices --exhibiting a linear decrease with lowering $T$
at first, and a tendency to saturate at lower $T$. The magnitude of
$V_T$ depends on the sample, and it is systematically smaller in
higher mobility devices (i.e., the experiments indicate that the
temperature dependence of $\mu$ correlates with the values of
$V_T$).

\begin{figure}[htbp]
\includegraphics[width=1\linewidth]{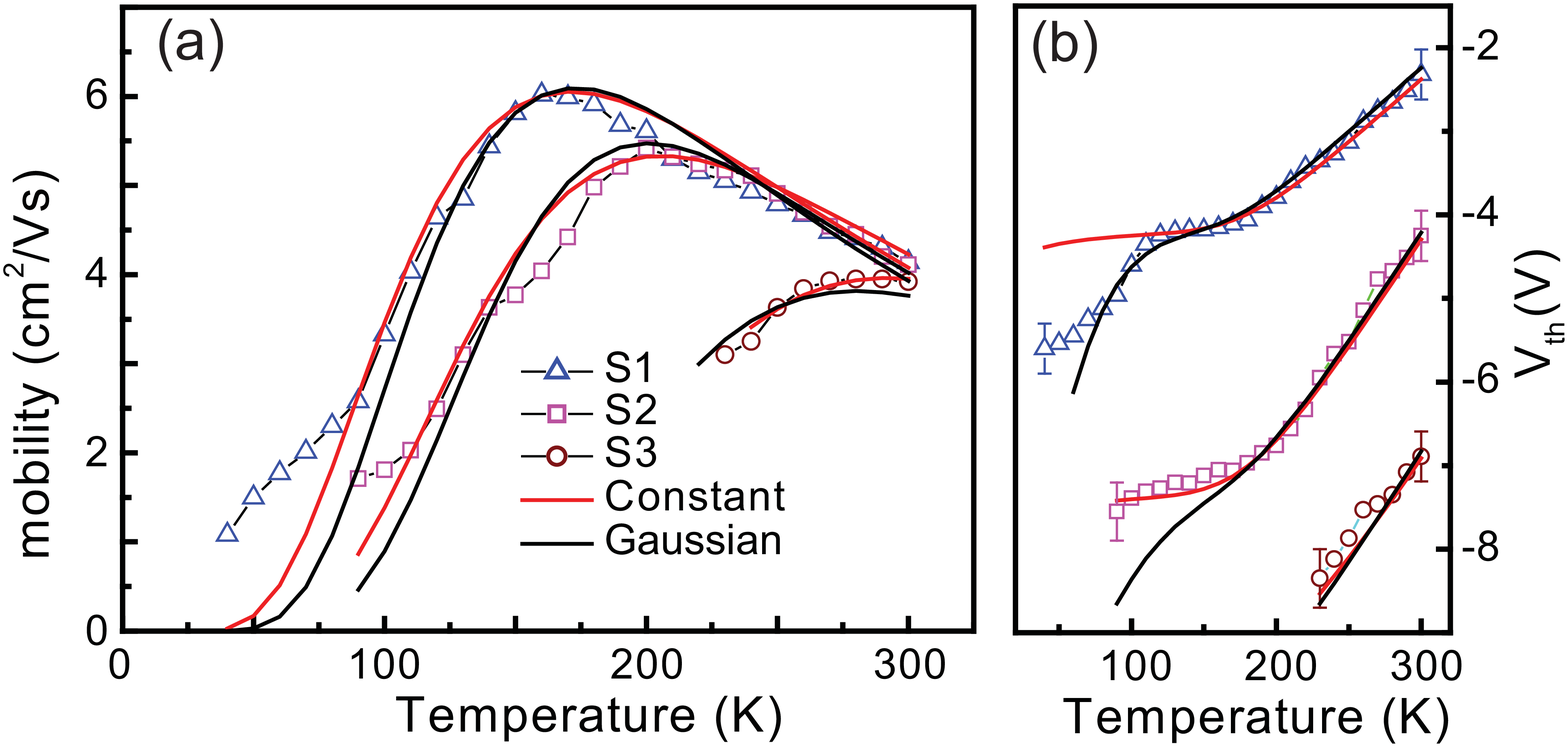}
  \caption {The symbols represent the  temperature dependent mobility (a) and
threshold voltage (b) measured on three different devices. The solid lines represents the result of our model, using the parameters listed in Table 1; the red and black lines are obtained using a constant and a gaussian density of shallow traps, respectively. }
  \label{FIG. 3}
\end{figure}

To analyze quantitatively the data we introduce a model based on the
assumption that transport of holes in TMTSF occurs in a band (with
density of  states $N_0$ for energy $E>0$), and that
disorder-induced trap states are present (for $E<0$)(see also \cite{Merlo}). These consist
of shallow traps (with density of states $N_S(E)$) and deep traps
located much deeper in energy (described by a density of states
$N_{D}(E)$). The mobility of carriers occupying states in the band
is the intrinsic mobility in the material, and, following existing
theories, is assumed to depend on temperature as $\mu_0(T)=\alpha
T^{-2}$\cite{Troisi, Hannewald, exponent}. Carriers occupying trap
states are assumed not to carry current.

In practice, we fix the shape of the density of states and calculate the position of the Fermi level $E_F$ (numerically) as a function of the total density of accumulated charge $n$ ($= C_i V_G$) through the relation:
\begin{eqnarray}
\label{Fermi distribution}
 n=&&\int_{0}^{+\infty}N_0\frac{1}{e^{\frac{E-E_F}{kT}}+1}dE \nonumber\\
 &&+\int_{-\infty}^{0}(N_S(E)+N_D(E))\frac{1}{e^{\frac{E-E_F}{kT}}+1}dE
\end{eqnarray}
that defines $E_F$. Having determined $E_F(n,T)$, it is straightforward to obtain the source-drain current $I_{SD}$ in terms of the density of charge carriers occupying states in the band ($n_C(n,T)=\int_{0}^{+\infty}N_0\frac{1}{e^{\frac{E-E_F}{kT}}+1}dE$) as:
\begin{equation}
I_{SD}(n=C_iV_G)= n_C(n,T)e \ \mu_0(T) \frac{W}{L}V_{SD},
\end{equation}
From Eq. 2 we extract all quantities that are measured in the experiments, by applying to the calculated $I_{SD}-V_{SD}$ curves the same procedures used for the measured ones. For instance, the mobility $\mu$ measured from the FET characteristics is simply given by $\mu = \mu_0(T) \frac{\partial n_c(n,T)}{\partial n}$ (which corresponds to $\mu \ \alpha \ \frac{\partial I_{SD}}{\partial V_G}$). For our calculations we use, for the shallow traps, a constant density of states ($N_S(E)= N_c =$ const, if $E_c < E <0$, and $N_S(E)=0$ otherwise), and a gaussian density of states ($N_S(E)= N_g \ e^{-(E/2E_g)^2}$), and compare the results obtained imposing that the total number of shallow traps $\int_{-\infty}^{0}N_S(E)dE$ is the same for the two distributions. For the deep traps we use a "square" distribution ($N_D(E)= N_D=$ constant if $E_D-\delta E_D/2 < E < E_D+\delta E_D/2$ and zero otherwise) to analyze the role of the trap depth ($E_D$) and of the distribution width ($\Delta E_D$). The density of states in the band at the surface of the TMTSF crystal -- $N_0 = 10^{15}$ cm$^{-2}$ eV$^{-1}$-- is estimated by taking one state per molecule distributed in energy over the bandwidth of the valence band ($\simeq 0.5$ eV \cite{Silinsh}).  The parameter $\alpha$ (in the expression $\mu_0(T)=\alpha T^{-2}$) represents an intrinsic property of TMTSF crystals, and it is taken to be the same for all samples.

A strategy similar to the one adopted here --i.e., "fixing" a density of states to analyze transport through FETs-- is used to describe transport in FETs based on amorphous materials\cite{Brutting}. In that case, however, transport is due to carriers occupying localized states, whose mobility is strongly energy dependent, and material inhomogeneity requires percolation effects to be taken into account\cite{Matters}. This significantly increases the complexity of the analysis, as well as the unambiguous extraction of parameters in the model. These problems are absent in high-quality single-crystal FETs. Only at low temperatures, such that most carriers occupy trap states, the model starts to deviate from the experiments, mainly because our assumption that carriers in the shallow traps do not carry any current is too restrictive. For this reason, we limit our analysis to a temperature range where a significant fraction of carriers populate states in the band. In this temperature range (between approximately 100 and 300 K for the best device) we find that the behavior of the temperature dependence of the mobility $\mu(T)$ measured at high carrier density curve depends only on the properties of the shallow traps (Fig. 3a), whereas the behavior of the $V_T(T)$ is determined only by the deep traps. This considerably simplifies the analysis.

\begin{figure}[htbp]
\includegraphics[width=1\linewidth]{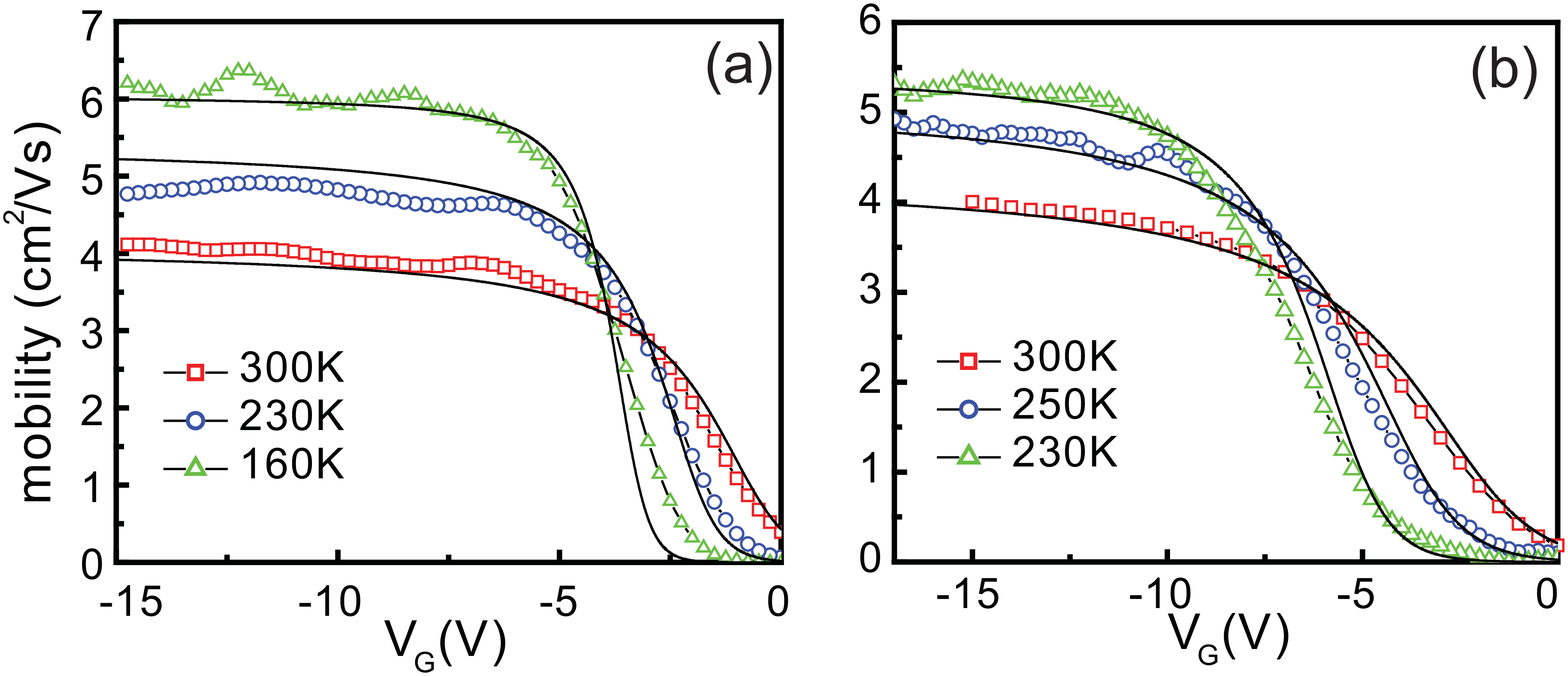}
  \caption {In panel (a) and (b) the symbols show the mobility extracted from FET measurements as a function of carrier density at different temperatures, on devices 1 and 2. The solid lines are calculated using the parameter values extracted from the fits shown in Fig. 3 (no other parameters need to be introduced). For device 3, a similar agreement between data and model is obtained.}
  \label{FIG. 4}
\end{figure}

Best fits to the temperature dependence of the mobility are shown in
Fig. 3(a), using both the gaussian and the constant distributions
of shallow traps. With both  distributions the agreement is
satisfactory for the temperature range discussed above\cite{exponential}. Table 1
summarizes the parameters extracted from the fits. Note that
the energies $E_c$ and $E_g$ satisfy the relation $E_c
= 3 E_g$ to a 10\% precision (or better) for all samples, which can
be easily understood since the integral of the gaussian trap
distribution between $E=3 E_g$ and $E=0$ gives 99\% of the shallow
trap states. Since different distributions lead to nearly
identical temperature dependencies of the mobility, the precise
energy distribution of the shallow traps cannot be extracted from
the analysis. On the contrary, the total density of shallow traps
and of their characteristic energy ($\simeq E_g$) are not sensitive to the
details of the model, and the values extracted have therefore physical meaning.

The microscopic analysis of the threshold voltage is meaningful,
because in our FETs the single-crystals are suspended on top of the
gate, which eliminates extrinsic effects due to phenomena taking
place in the gate insulator. We find that the curve $V_T(T)$ is
sensitive to both the energy and the width of the deep trap
distribution. Specifically, very deep traps (whose energy is larger
than roughly 0.5 eV) contribute to the threshold voltage, but do not
cause a shift in the temperature range considered.  In this same
range, a sizable contribution to the threshold voltage shift
originates from traps with energy between 100 and 400 meV,
approximately, with the largest contribution coming from traps
between 150-200 meV. This fact --that deep traps at different
energies affect differently  $V_T$ and its shift with temperature--
makes it possible to extract information on the energy distribution
of deep traps from the experimental $V_T(T)$ curves. With
the parameter values listed in Table 1, the model leads to an
excellent quantitative agreement with the data (again, below approximately 100 K,
where the majority of carriers occupies shallow traps, the results depend on the
details of the shallow trap distribution). Note that the center
of the deep trap distributions is approximately the same in the three devices. This
is expected if deep traps originate from a well-defined chemical
impurity hosted in the crystal. Indeed, the values of $E_D$ and of
$\delta E_D$ extracted from our analysis are in the same range energies and widths
of other deep traps recently found in different molecular
materials\cite{deeptraps}.

\begin{table}
\begin{ruledtabular}
\begin{tabular}{llll}
  & Sample1 &Sample2 &Sample3\\
\hline
\\
N$_{C}$(cm$^{-2}$\ eV$^{-1}$) & 1.97*10$^{14}$ & 8.5*10$^{13}$ & 9.3*10$^{12}$ \\
E$_{C}$(meV)&38.5 & 54.1 & 111 \\
\hline
\\
N$_{g}$(cm$^{-2}$\ eV$^{-1}$)&3.04*10$^{14}$ & 1.36*10$^{14}$ & 1.42*10$^{13}$ \\
E$_{g}$(meV)&14.1 &19.1 & 40.8  \\
\hline
\\
$\int_{-\infty}^{0}N_S(E)dE$ (cm$^{-2}$)&7.6*10$^{12}$ &4.6*10$^{12}$ & 1.03*10$^{12}$ \\
\hline \hline
\\
N$_{D}$(cm$^{-2}$ \ eV$^{-1}$)&6*10$^{10}$ & 1.08*10$^{11}$ & 1.08*10$^{11}$ \\
E$_{D}$(meV)&225 & 235 & 270 \\
$\delta$E$_{D}$(meV) & 95 & 95 & 130\\
\hline
\\
$\Delta\Phi$(meV) &10.3 & 13.7 & 16.1 \\
\end{tabular}
\end{ruledtabular}
\caption {Values of  quantities discussed in the text. The total number of shallow
traps $\int_{-\infty}^{0}N_S(E)dE$ is fixed to be the same for both the constant and the gaussian
energy distributions.  }
\end{table}

With the model reproducing successfully the experimental
observations, it is important to critically ask whether the
quantitative agreement is not simply due to the introduction of a
sufficient number of free parameters. To show that this is not the
case, we have analyzed the full dependence of the
mobility on carrier density, at different temperature values. This
comparison provides a stringent test of the validity of the model
and of its consistency, because all parameters are fixed by the analysis discussed
above, and no other free parameters can be introduced. Fig. 4a and b
show that the curves calculated with the model are in excellent
quantitative agreement with the data, as long as the temperature is
such that enough charge carriers are present in the band. This
result confirms the internal consistency of our analysis, and
illustrates that the model does describe well the interplay between
band-like transport and trapping as a function of carrier density.

Having established the internal consistency of our analysis, we
address the origin of the observed correlation between the
temperature dependence of mobility and of threshold voltage. Since
the temperature dependence of the mobility is determined only by the
shallow traps and that of the threshold voltage only by the deep
traps, this correlation implies that a physical mechanism exists,
that links the origin of the two kinds of traps. We propose that
this mechanism is due to simple electrostatics.  Individual charge
carriers frozen in the deep traps generate random electrostatic
potential fluctuations $\Delta \Phi$, and the resulting
electrostatic potential landscape causes regions of lower energy in
which a large density of charge carriers can be trapped. The
characteristic energy scale associated to the potential fluctuations
can be estimated as $\Delta \Phi = \frac{1}{4 \pi \epsilon
\epsilon_0} \frac{e}{<r>}$, where $<r>$ is the average distance
between deep traps and $\epsilon$ is the average of the relative
dielectric constant of the TMTSF crystal and of vacuum. By
estimating the magnitude of $<r>$ from the total density of deep
traps extracted from the fit of $V_T(T)$, we obtain the values of
$\Delta \Phi$ reported in Table 1, whose magnitude compares well
with the characteristic energy scale $E_g$ of the shallow
traps. This scenario explains why the density of states of shallow
traps at $E\simeq 0$ is comparable to the density of states in the
band itself. In fact, shallow traps consist of states in the band
whose energy is lowered by the local, slowly varying electrostatic
potential. It  also explains why carriers in shallow traps
contribute significantly to the conductivity (as we noticed
earlier), since these carriers are not tightly bound to a defect and
can move over rather large distances.

In conclusion, we have developed a well-defined framework to analyze
the interplay between intrinsic band-like transport and trapping in
organic single-crystal transistors and used it to describe the
behavior of newly developed TMTSF FETs. The model permits to understand in detail the physical
origin of the different aspects of the device characteristics, and to extract microscopic
parameters reliably. This combination of experimental investigation on high-quality devices
and systematic quantitative analysis is important to reach a true
understanding of molecular semiconductors at finite
density of charge carriers.

We are grateful to S. Fratini for discussions and comments, and we
acknowledge financial support from NanoNed, the Swiss NCCR MaNEP, NEDO, and FCT.


\begin{references}
\bibitem{exponent} Presently, our analysis cannot discriminate between small differences in the value of the exponent, which depend on details of the theoretical models.
\bibitem{exponential} We have obtained similar and consistent results also with
an exponential distribution. With such a distribution, however, the threshold voltage below 100-150 K
is suppressed much more rapidly than with the constant or gaussian distributions.
\end{references}
\end{document}